\documentclass[aps,prd,nofootinbib,amsmath,amssymb,superscriptaddress,tightenlines,11pt]{revtex4}
\usepackage{graphicx}
\usepackage{dcolumn}
\usepackage{bm}
\usepackage{amssymb}
\usepackage{latexsym}
\usepackage{booktabs}
\usepackage[colorlinks, linkcolor=blue, citecolor=blue, urlcolor=blue]{hyperref}

\newcommand{\be}{\begin{equation}}
\newcommand{\ee}{\end{equation}}
\newcommand{\bq}{\begin{eqnarray}}
\newcommand{\eq}{\end{eqnarray}}

\begin{document}

\title{Fitting to data of superluminal neutrinos with phenomenological scenarios}

\author{Li-Ang Zhao}
\affiliation{Department of Physics, College of Sciences,
Northeastern University, Shenyang 110004, China}
\author{Xin Zhang}
\email{zhangxin@mail.neu.edu.cn} \affiliation{Department of Physics,
College of Sciences, Northeastern University, Shenyang 110004,
China} \affiliation{Center for High Energy Physics, Peking
University, Beijing 100080, China}

\begin{abstract}
We test several phenomenological scenarios of faster-than-light
neutrinos, by fitting to the experimental data from OPERA, MINOS and
Fermilab79. Our purpose is to see, from the perspective of the
current data, whether or not the speed of the superluminal neutrino
depends on its energy. We show that the Coleman-Glashow scenario in
which the velocity of the neutrino is free of the energy fits the
data best. However, the result of SN1987A cannot be explained by
this model. We find that a power-law model with the power close to
zero can simultaneously explain the results of SN1987A and
OPERA+MINOS+Fermilab79.
\end{abstract}

\maketitle

The OPERA collaboration reported the superluminal muon neutrino
$(\nu_\mu)$ data \cite{1109.4897}, recently. The neutrinos with
average energy $\sim$ 17 GeV arrive earlier by
\begin{equation}
\delta t=(60.7\pm6.9~(\text{stat.})\pm7.4~(\text{sys.}))\;{\rm ns}
\end{equation}
than photons, from CERN to Gran Sasso Laboratory with distance about
730 km. This indicates that neutrinos are superluminal, with an
excess of the speed than light (in vacuum)
\begin{equation}
v-1=(2.48\pm0.28~(\text{stat.})\pm0.30~(\text{sys.}))\times10^{-5},
\end{equation}
with significance level of 6$\sigma$. Note that throughout the paper
we use the natural units with $c=1$. It was also found by OPERA that
the velocity difference $v-1$ is almost independent of energy, by
dividing the events into two groups with energies below or above 20
GeV: $v-1=(2.16\pm0.76~(\text{stat.})\pm0.30~(\text{sys.}))$ for
$\langle E\rangle=13.9$ GeV and
$v-1=(2.74\pm0.74~(\text{stat.})\pm0.30~(\text{sys.}))$ for $\langle
E\rangle=42.9$ GeV.

It should also be pointed out that earlier experiments ever obtained
similar results of superluminal neutrinos, though with lower
significance. For example, the MINOS experiment at Fermilab in 2007
\cite{0706.0437} and even earlier experiments at Fermilab in 1979
\cite{1979} reported the results of $v-1\sim 10^{-5}$, similar to
that of OPERA. These results are summarized in Table
\ref{table:data}. However, the supernova neutrinos from SN1987A
place a stringent velocity constraint, $v-1<2\times 10^{-9}$, for
electron neutrinos with energies from 5 to 40 MeV, seemingly
inconsistent with the superluminal neutrino result of OPERA.

At first glance, the result of that the neutrino travels faster than
light is too shocking, and many people instinctively reject such a
result. Indeed, there might be some systematic errors or some other
unknown factors in the experiment that have not been taken into
account. For example, it was suspected that there may be some
serious problems in the measurement of time and distance in the
experiment \cite{gps}. And, in the theoretical respect, Cohen and
Glashow \cite{CG} argued that superluminal neutrinos with high
energies would lose energy rapidly via bremsstrahlung of
electron-positron pairs, causing the beam to be depleted of higher
energy neutrinos, and so they refuted the superluminal
interpretation of the OPERA result. However, on the other hand, if
the nature of neutrinos is indeed superluminal, then the modern
physics would be revised enormously. Thus, before the independent
experiments, such as BOREXINO, ICARUS, and MINOS, confirm (or
negate) the OPERA result, it is a rational way for us to suppose
that the data and conclusion of OPERA are right and try to look for
the possible clues of the new revolution of physics via the
superluminal neutrino experiments.

In recent days, many theoretical explanations for the superluminal
neutrinos have been put forward. For example, it has been suggested
that such a phenomenon might originate from a violation of Lorentz
invariance
\cite{AmelinoCamelia:2011dx,Li:2011ue,Lingli:2011yn,Bi:2011nd}.
Nevertheless, it is hard to understand why the Lorentz violation
happens at such low energies, far below the energy scale of quantum
gravity. Tachyonic neutrino scenario is another attractive
explanation. However, this scenario is clearly disfavored by the
data \cite{AmelinoCamelia:2011dx}. Besides, there are also many
other possible mechanisms, such as, extra-dimension model
\cite{Gubser:2011mp}, neutrino dark energy model
\cite{Ciuffoli:2011ji}, and so on.

One important point should be clarified, in our opinion, before we
make deeper theoretical analysis: whether or not the superluminal
phenomenon of neutrinos is energy-dependent? In fact, it was
proposed by Li and Wang \cite{Li:2011ue} that the Lorentz violation
might be energy-dependent, only occurring in a window of particular
energies (including $3-200$ GeV), thus reconciling the results of
OPERA and SN1987A. In addition, Amelino-Camelia et al.
\cite{AmelinoCamelia:2011dx} also tackled the issue of the
energy-dependence of the speed of neutrinos. In
\cite{AmelinoCamelia:2011dx} several cases of violation or
deformation of Lorentz invariance were considered, and the authors
showed that the Coleman-Glashow (CG) scenario and the Doubly Special
Relativity (DSR) scenario with linear-dependence are more favored,
while the tachyon scenario and DSR model with quadratic-dependence
are clearly disfavored, by the data. Of course, it seems that the
data-favored models are inconsistent with the result of SN1987A. In
this paper, we will make a more detailed analysis of current data
with some phenomenological scenarios. We will show whether the
superluminal phenomenon of neutrinos is energy-dependent, according
to the current experimental data, and which phenomenological
scenario can fit the data well and be consistent with the result of
SN1987A at the same time.

\begin{figure*}[htbp]
\centering
\begin{center}
$\begin{array}{c@{\hspace{0.2in}}c} \multicolumn{1}{l}{\mbox{}} &
\multicolumn{1}{l}{\mbox{}} \\
\includegraphics[scale=0.3]{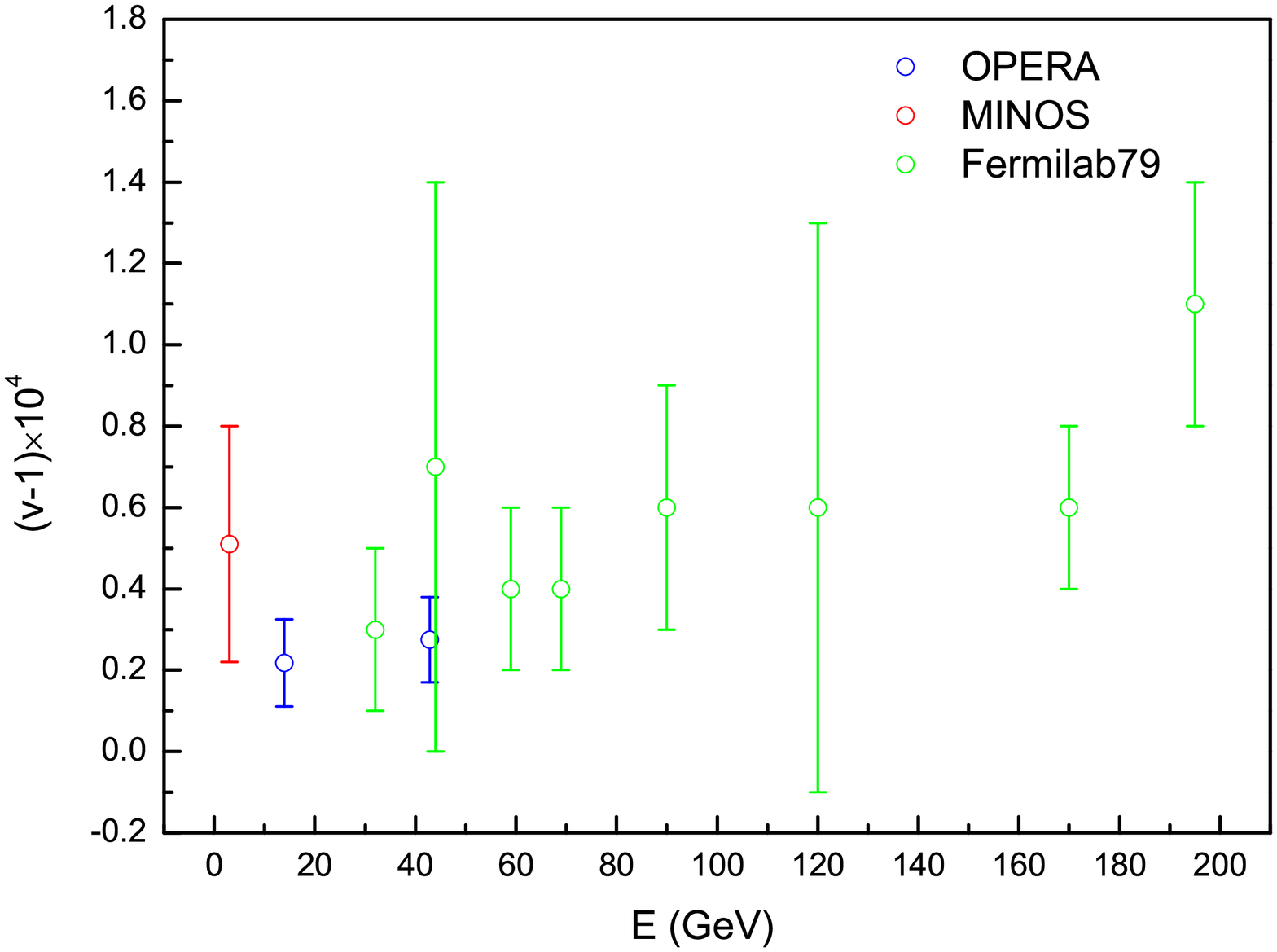} &\includegraphics[scale=0.3]{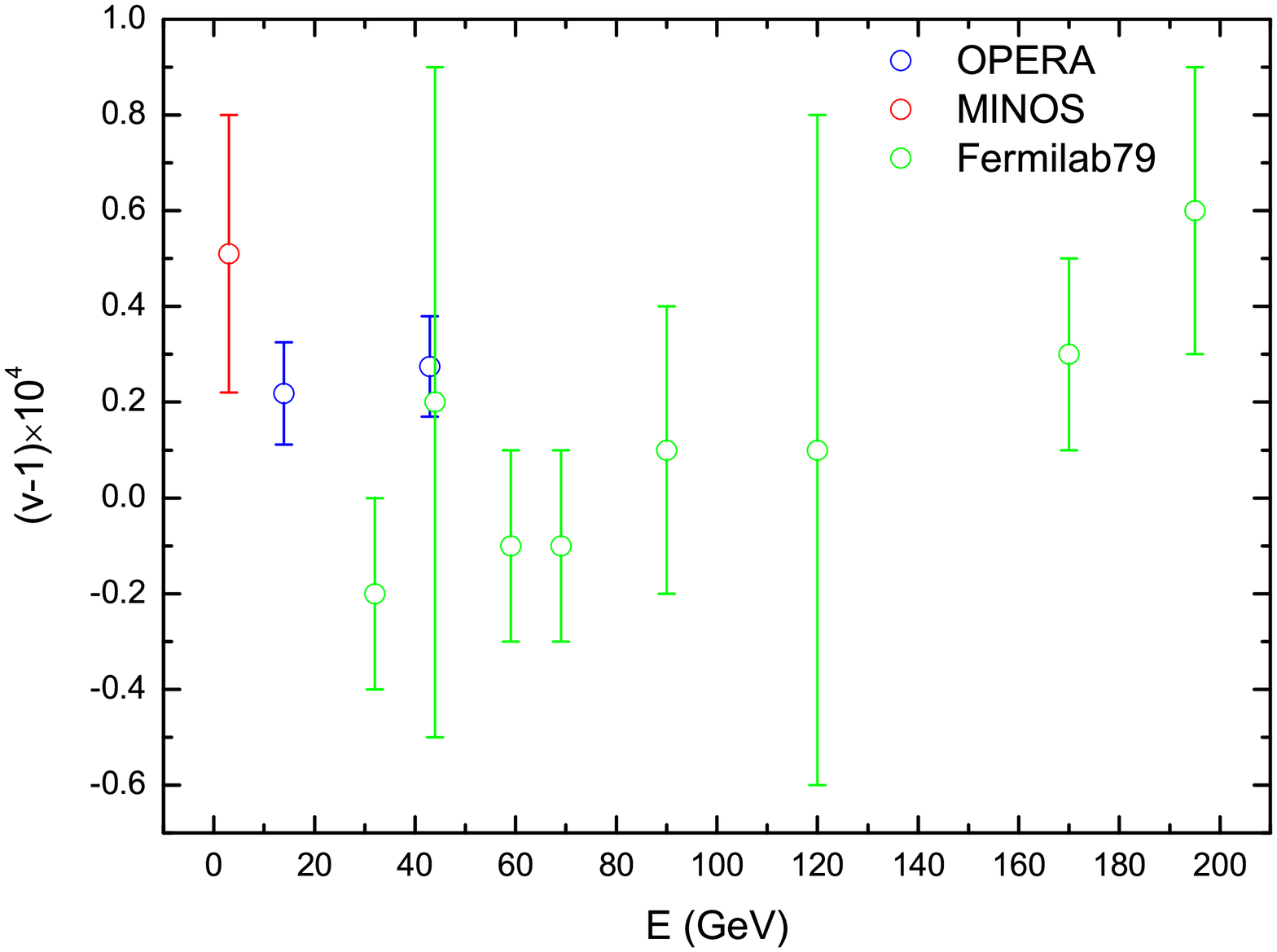} \\
%\mbox{(a)} & \mbox{(b)}
\end{array}$
\end{center}
\caption[]{\small \label{fig:data}The superluminal neutrino data
from OPERA, MINOS and Fermilab79. The left panel shows the raw data,
and the right panel shows the data corrected by
$b_{1979}=0.5\times10^{-4}$ for Fermilab79.}
\end{figure*}

We show the data of OPERA+MINOS+Fermilab79 in the left panel of
Fig.~\ref{fig:data}. As discussed in
\cite{1979,AmelinoCamelia:2011dx}, when using the data of
Fermilab79, a bias correction, downward shift of data with
$b_{1979}=0.5\times10^{-4}$, should be considered. The data with the
above correction (for Fermilab79) are shown in the right panel of
Fig.~\ref{fig:data}.

\begin{table*}
\caption{The summary of superluminal neutrinos from OPERA, MINOS,
and Fermilab79.}\label{table:data}
\begin{tabular}{|c|c|c|c|c|}
\hline
Experiment & Velocity constraint & Energy range & Reference\\
\hline OPERA & $v-1=(2.48\pm0.58)\times10^{-5}$ (6.0~$\sigma$)
& $\sim17$~GeV&\cite{1109.4897}\\
\hline MINOS & $v-1=(5.1\pm2.9)\times10^{-5}$ (68\% CL)
& $\sim3$~GeV &\cite{0706.0437} \\
\hline Fermilab79  & $|v-1|<4\times10^{-5}$ (95\% CL) & $30$ to
$200$~GeV &\cite{1979}\\
\hline
\end{tabular}
\end{table*}

In the ultra-relativistic regime, the Einstein's mass-energy
relation gives the following description of the dependence of speed
on energy:
\begin{equation}\label{v-1}
v-1=-\frac{1}{2} \frac{m^2}{E^2}.
\end{equation}
Obviously, in the theory of relativity a normal particle with real
mass $m$ can never travel faster than light in the vacuum. Now, let
us estimate the order of magnitude of the velocity difference for
the neutrino by using Eq.~(\ref{v-1}). If the mass of neutrino is
supposed to be $m\sim$ eV and the neutrino energy $E$ under
consideration ranges from 3 GeV to 200 GeV, we obtain
$m^2/2E^2\sim10^{-18}-10^{-22}$. So, even if we consider a tachyon
model in which the neutrino has an imaginary mass, ${\cal
M}^2=-m^2$, the theoretical value obtained is many order of
magnitude less than the experimental result, ${\cal O}(10^{-5})$.

Supposing that there is a theoretical mechanism that can explain the
experiments such as OPERA, the above relation can be modified,
phenomenologically, as
\begin{equation}\label{v-1:f}
v(E)-1=-\frac{1}{2} \frac{m^2}{E^2}+f(E)\approx f(E),
\end{equation}
where the form of $f(E)$ is determined by the underlying theoretical
mechanism that is unknown for us now but can be modeled
phenomenologically. We can view that $f(E)$ comes from a Lorentz
violation, though we cannot understand why the energy scale of such
a Lorentz violation is so low. Alternatively, we can even consider a
mass-running tachyonic neutrino scenario with mass ${\cal
M}(E)^2=-m(E)^2=2E^2f(E)$.

In what follows, we take a totally phenomenological perspective,
assuming the possible forms of $f(E)$. First, we consider the
possible form inspired by Coleman and
Glashow~\cite{Coleman:1997xq,Coleman:1998ti},
\begin{equation}\label{f1}
f(E)=\delta,
\end{equation}
where $\delta$ is a constant parameter. Also, we follow
Amelino-Camelia et al. \cite{AmelinoCamelia:2011dx}, taking the
DSR-type cases, $f(E)=\ell_1E$ and $f(E)=\ell_2^2E^2$. We,
furthermore, consider a more general form, a combination of the
above possible cases,
\begin{equation}\label{a2b1d0}
f(E)=\alpha E^2+\beta E+\delta.
\end{equation}
We will use the superluminal neutrino data currently available,
namely, the combination of OPERA, MINOS and Fermilab79, to tell us
which scenario works well. To fit the data, we use a $\chi^2$
statistic,
\begin{equation}
\chi^2(\bm\theta)=\sum\limits_{i=1}^{11}{(v^{\rm
mod}(E_i;\bm\theta)-v^{\rm exp}(E_i))^2\over\sigma(E_i)^2},
\end{equation}
where $\sigma(E_i)$ is the 1$\sigma$ error for each datum of
$v(E_i)$, and $\bm\theta$ denotes the model parameters. By
minimizing $\chi^2$, we can find the best-fitted parameters of the
models, and further obtain the probability contours in the
parameter-planes.

\begin{figure*}[htbp]
\centering
\begin{center}
$\begin{array}{c@{\hspace{0.2in}}c} \multicolumn{1}{l}{\mbox{}} &
\multicolumn{1}{l}{\mbox{}} \\
\includegraphics[scale=0.3]{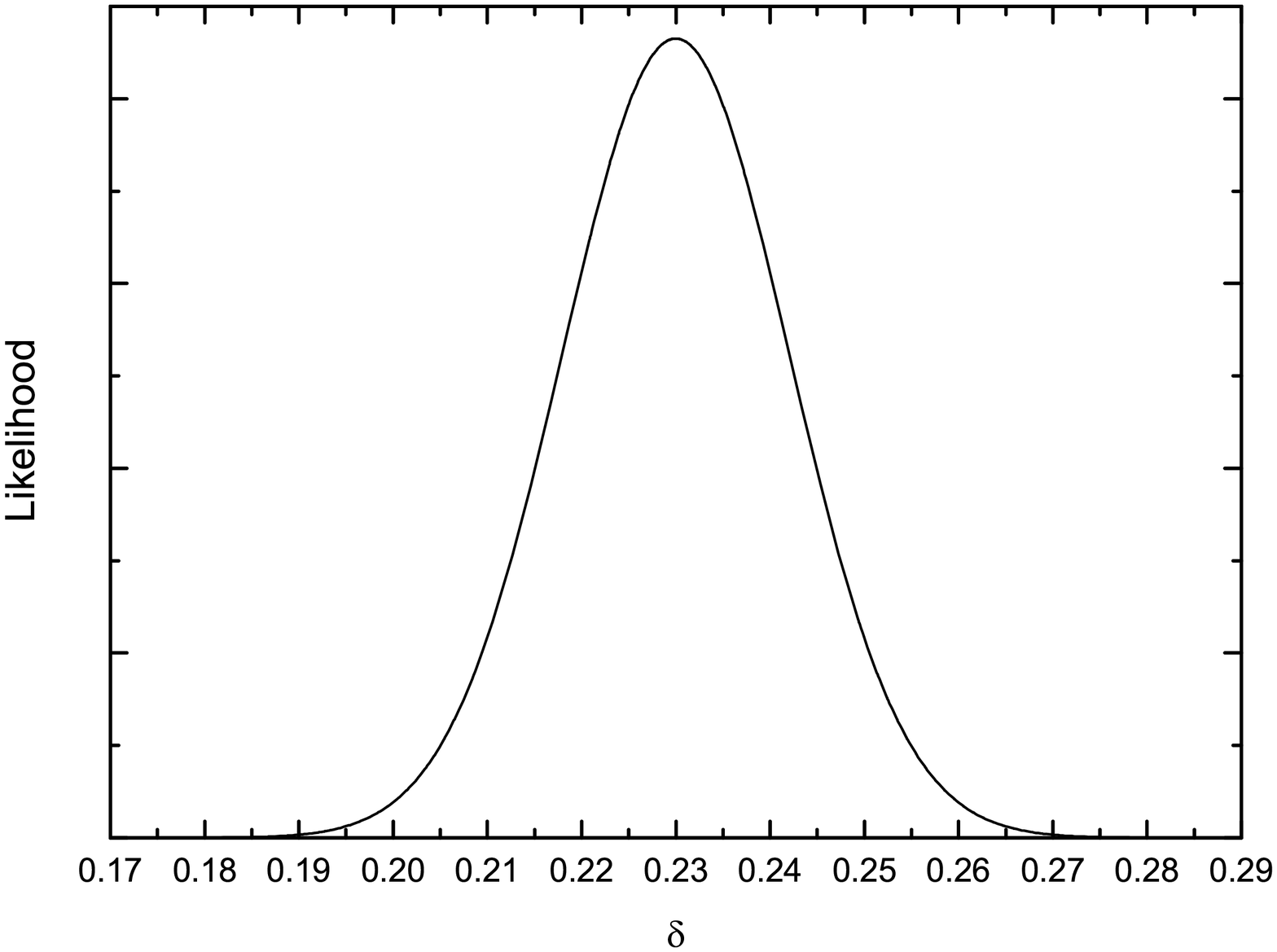} &\includegraphics[scale=0.3]{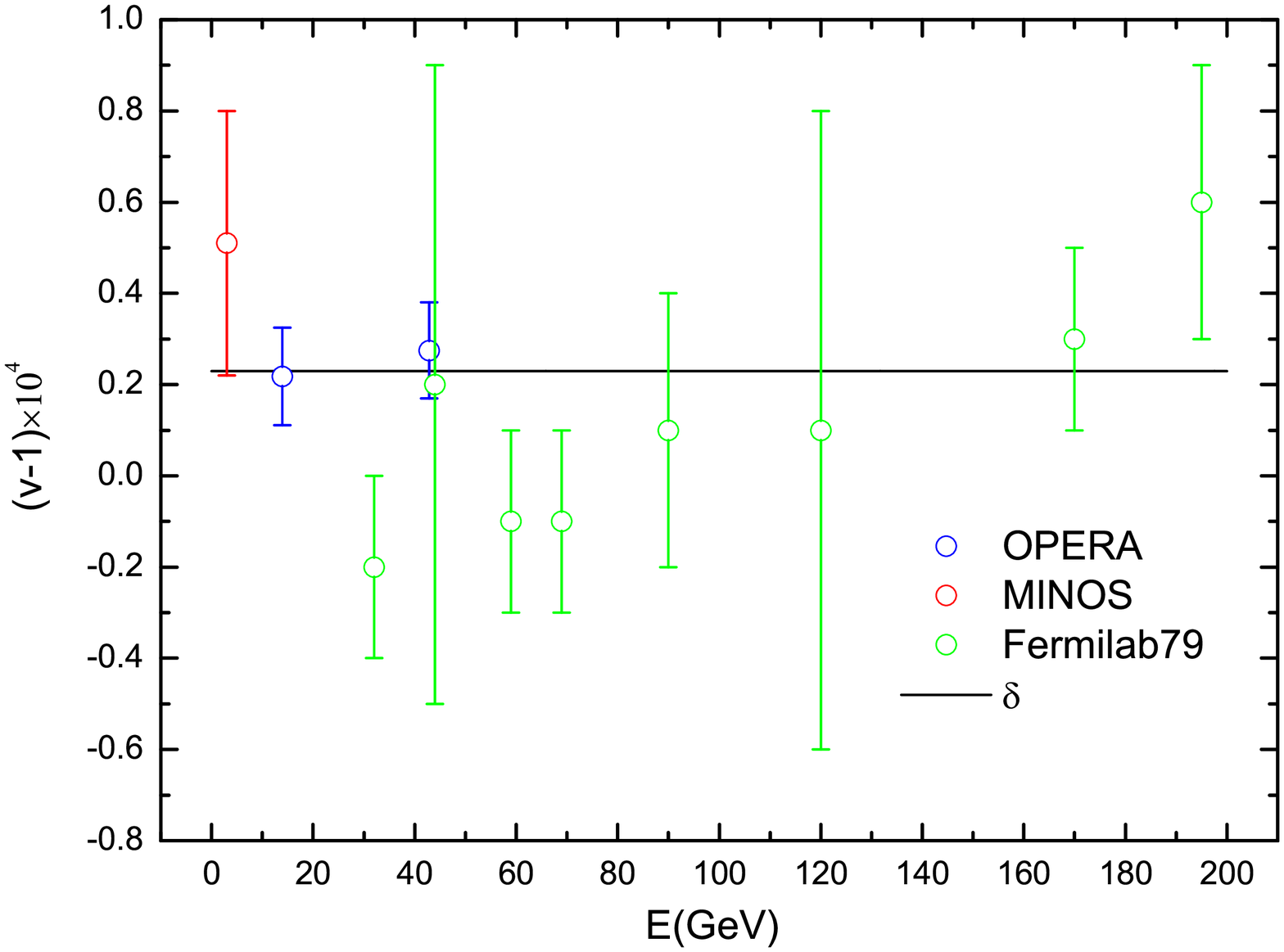} \\
%\mbox{(a)} & \mbox{(b)}
\end{array}$
\end{center}
\caption[]{\small \label{fig:CG}Likelihood distribution of parameter
$\delta$ for the Coleman-Glashow scenario, $f(E)=\delta$.}
\end{figure*}

\begin{figure*}[htbp]
\centering
\begin{center}
$\begin{array}{c@{\hspace{0.2in}}c} \multicolumn{1}{l}{\mbox{}} &
\multicolumn{1}{l}{\mbox{}} \\
\includegraphics[scale=0.3]{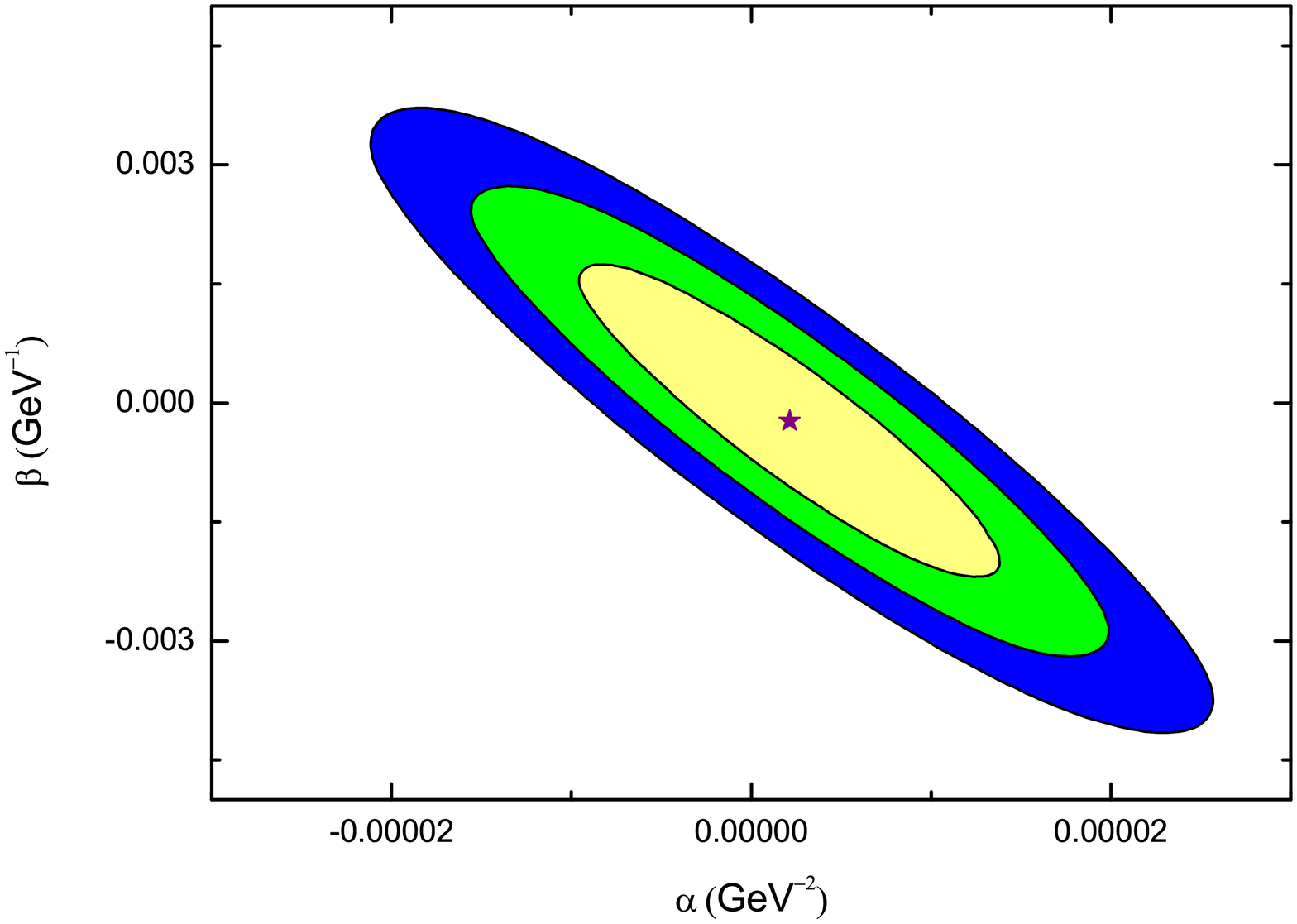} &\includegraphics[scale=0.3]{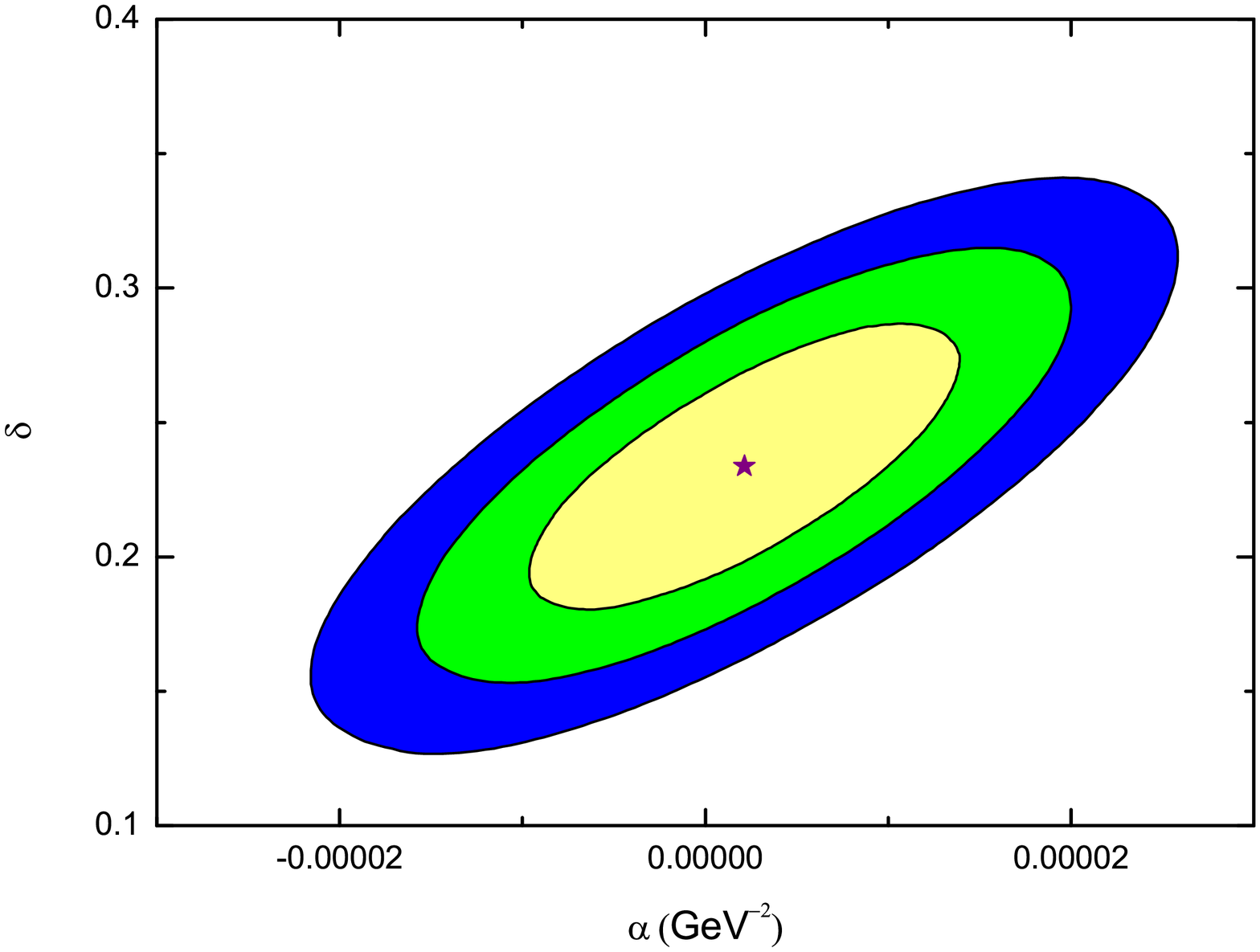} \\
%\mbox{(a)} & \mbox{(b)}
\end{array}$
$\begin{array}{c@{\hspace{0.2in}}c} \multicolumn{1}{l}{\mbox{}} &
\multicolumn{1}{l}{\mbox{}} \\
\includegraphics[scale=0.3]{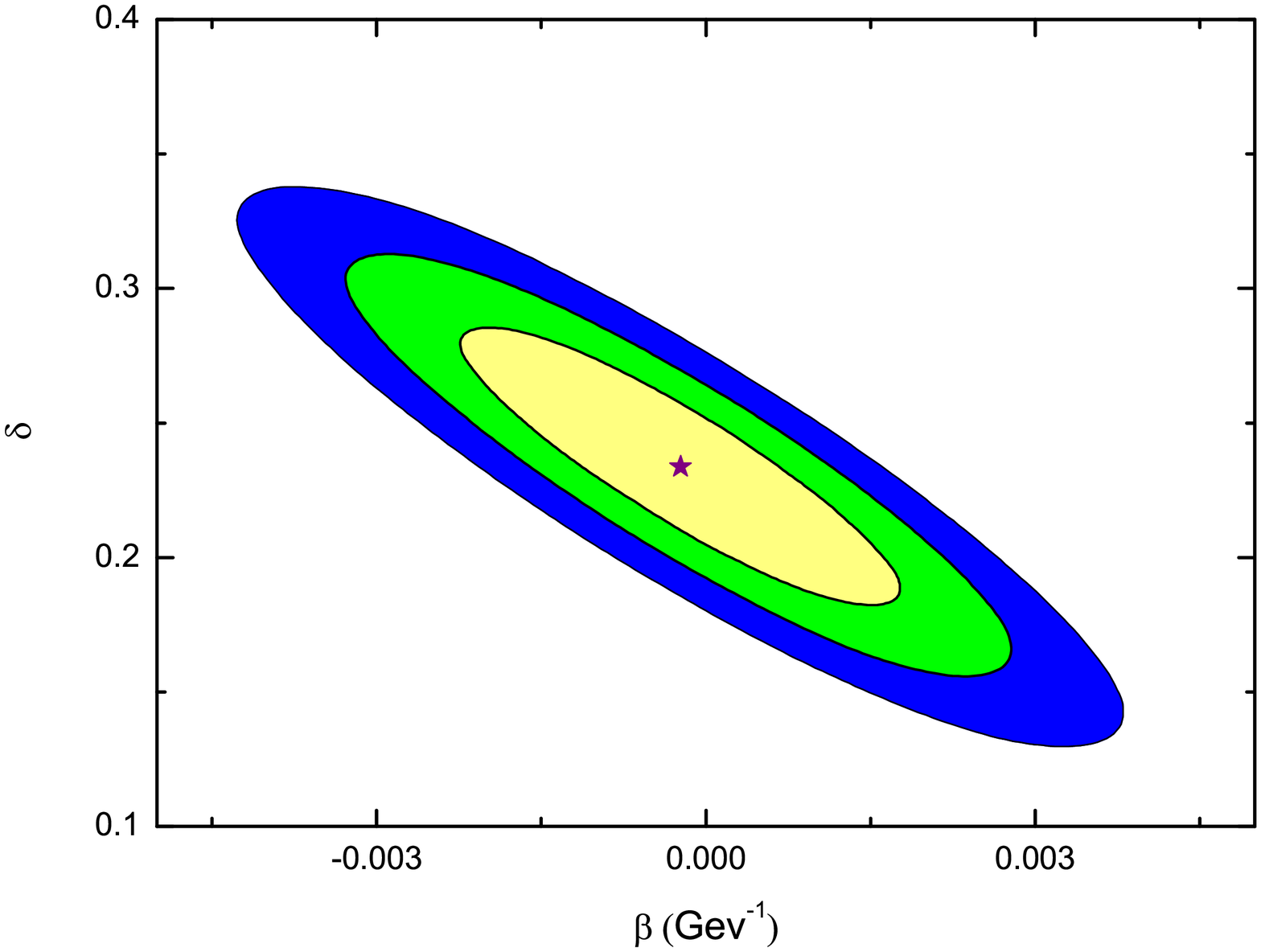} &\includegraphics[scale=0.3]{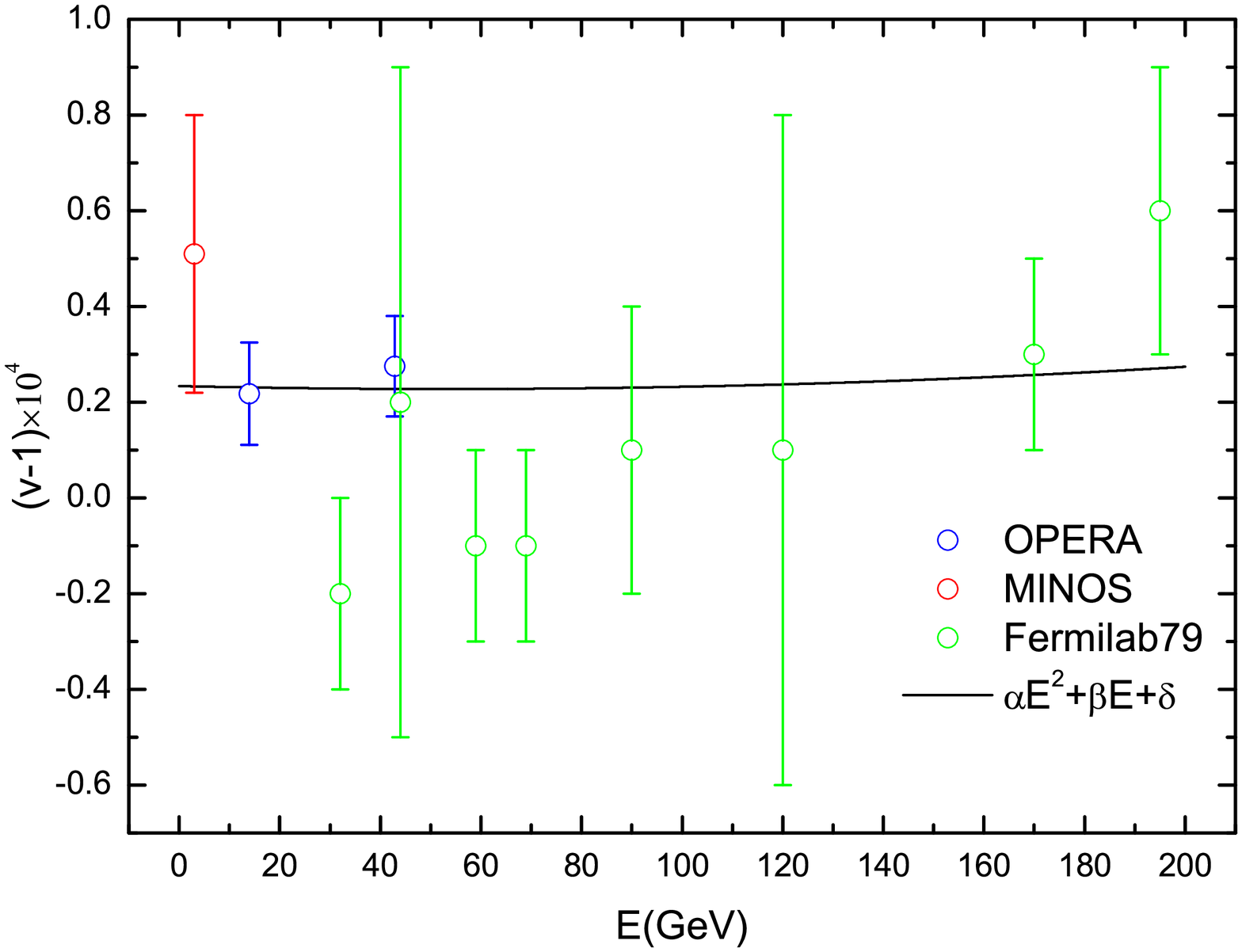} \\
%\mbox{(c)} & \mbox{(d)}
\end{array}$
\end{center}
\caption[]{\small \label{fig:general}Probability contours at 68.3\%,
95.4\% and 99.7\% confidence level in $\alpha-\beta$,
$\alpha-\delta$ and $\beta-\delta$ planes for the scenario,
$f(E)=\alpha E^2+\beta E+\delta$.}
\end{figure*}

For the CG scenario, we get $\chi^2_{\rm min}=56.94$. The DSR-type
cases, $f(E)=\ell_1E$ and $f(E)=\ell_2^2E^2$, are both clearly
disfavored by the data, leading to much bigger $\chi^2_{\rm min}$,
being 195.46 and 368.18, respectively. The combination form,
$f(E)=\alpha E^2+\beta E+\delta$, gives the least $\chi^2_{\rm
min}$, 56.77, among these scenarios. However, it is unwise to use
the $\chi^2$ statistic alone to compare competing models since the
number of parameters is different for the models. In general, a
model with more parameters tends to give a lower $\chi^2_{\rm min}$,
so instead one may employ the information criteria (IC) to assess
different models. In this paper, we use the BIC (Bayesian
information criterion) and AIC (Akaike information criterion) as
model selection criteria, defined respectively as
\begin{equation}\label{ic}
{\rm BIC}=\chi^2_{\rm min}+k\ln N,~~~{\rm AIC}=\chi^2_{\rm min}+2k,
\end{equation}
where $k$ is the number of parameters, and $N$ is the data points
used in the fit. Note that the absolute value of the criterion is
not of interest, only the relative value between different models,
$\Delta{\rm BIC}$ or $\Delta{\rm AIC}$, is useful. For the details
about the BIC and AIC, especially their applications in a
cosmological context, see, e.g., \cite{Liddle:2004nh,Li:2009jx}.
According to the fitting result, we see that the CG scenario has the
minimal values of BIC and AIC. The scenario of $f(E)=\alpha
E^2+\beta E+\delta$ yields $\Delta{\rm BIC}=4.62$ and $\Delta{\rm
AIC}=3.83$, relative to the CG case; see Table~\ref{result}. So,
according to the principle of Occam¡¯s razor, ``entities must not be
multiplied beyond necessity,'' the CG scenario, $f(E)=\delta$, is
preferred by the data. Figure~\ref{fig:CG} shows the fit result of
the CG scenario, with the left panel the one-dimensional likelihood
distribution of the parameter $\delta$, and the right panel the
best-fitted case of $v(E)$ comparing to the data.
Figure~\ref{fig:general} shows the fit result of the scenario of
$f(E)=\alpha E^2+\beta E+\delta$. From this figure, we can see that,
though the parameters $\alpha$ and $\beta$ are both around zero at
the best fit, there are some evident degeneracies between the
parameters. From the probability contours in the parameter-planes,
we find that, $\alpha$ and $\beta$ are anti-correlated, $\beta$ and
$\delta$ are also anti-correlated, and so $\alpha$ and $\delta$ are
in positive correlation. The best-fitted $v(E)$ curve is similar to
a horizontal line, but with a slight slope in the right-hand region.

\begin{table*}
\tabcolsep 5mm \caption{The fit results of the phenomenological
scenarios.}\label{result}
\begin{center}
\begin{tabular}{|c|c|c|c|c|}
\hline $f(E)$ scenario & $\chi^2_{\rm min}$ & $k$ & $\Delta$BIC & $\Delta$AIC \\
\hline
$\delta$ & 56.94 & 1 &0&0\\
\hline
$\ell_1 E $ & 195.46 & 1 &138.52&138.52\\
\hline
$\ell_2^2 E^2$ & 368.18 & 1 &311.24&311.24\\
\hline
$\alpha E^2+\beta E+\delta$ & 56.77 & 3 &4.62&3.83\\
\hline
$\xi (E/$GeV$)^\lambda$& 56.78 & 2 &2.24&1.84\\
\hline
\end{tabular}
\end{center}
\end{table*}

\begin{figure*}[htbp]
\centering
\begin{center}
$\begin{array}{c@{\hspace{0.2in}}c} \multicolumn{1}{l}{\mbox{}} &
\multicolumn{1}{l}{\mbox{}} \\
\includegraphics[scale=0.3]{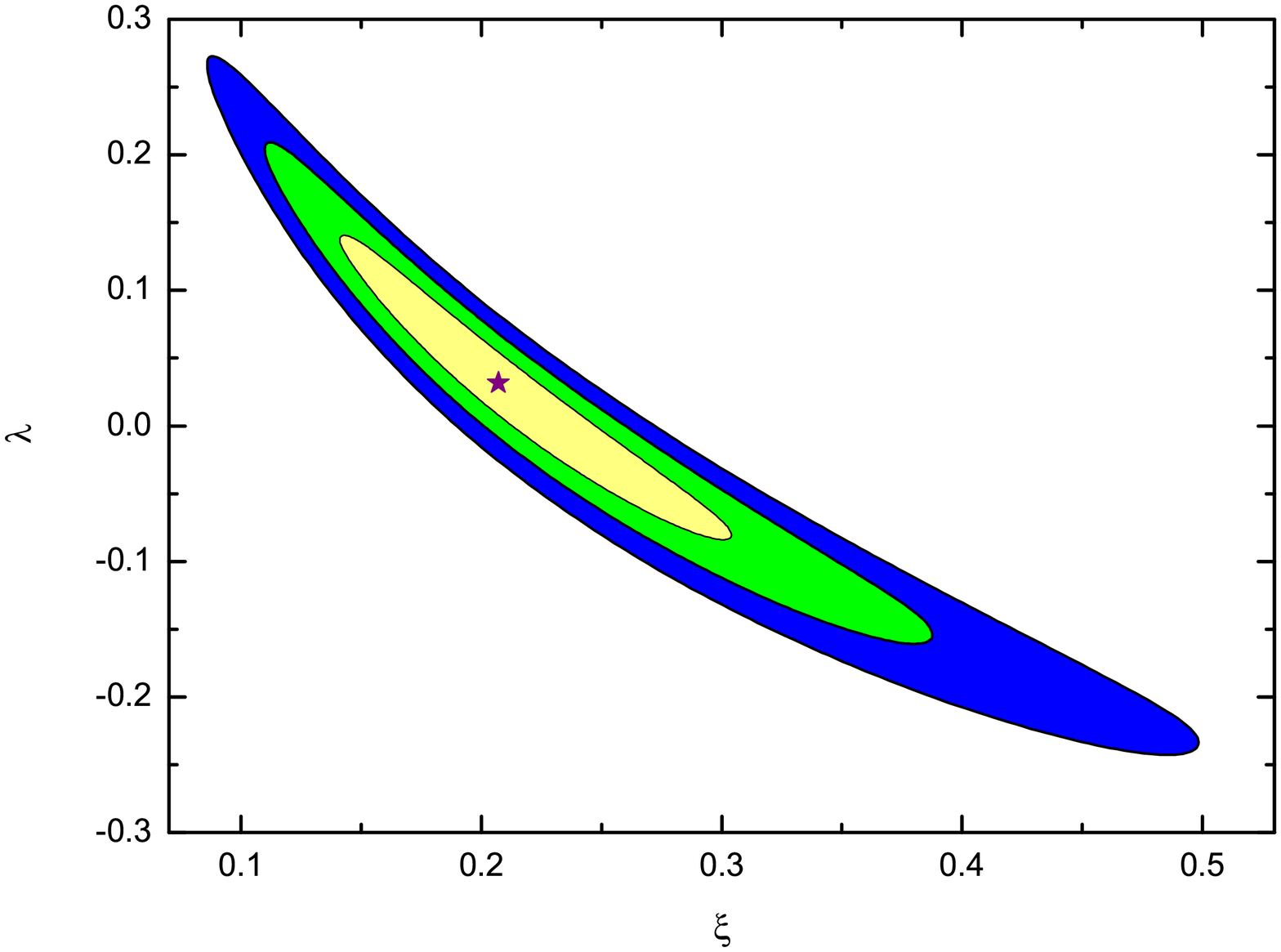} &\includegraphics[scale=0.3]{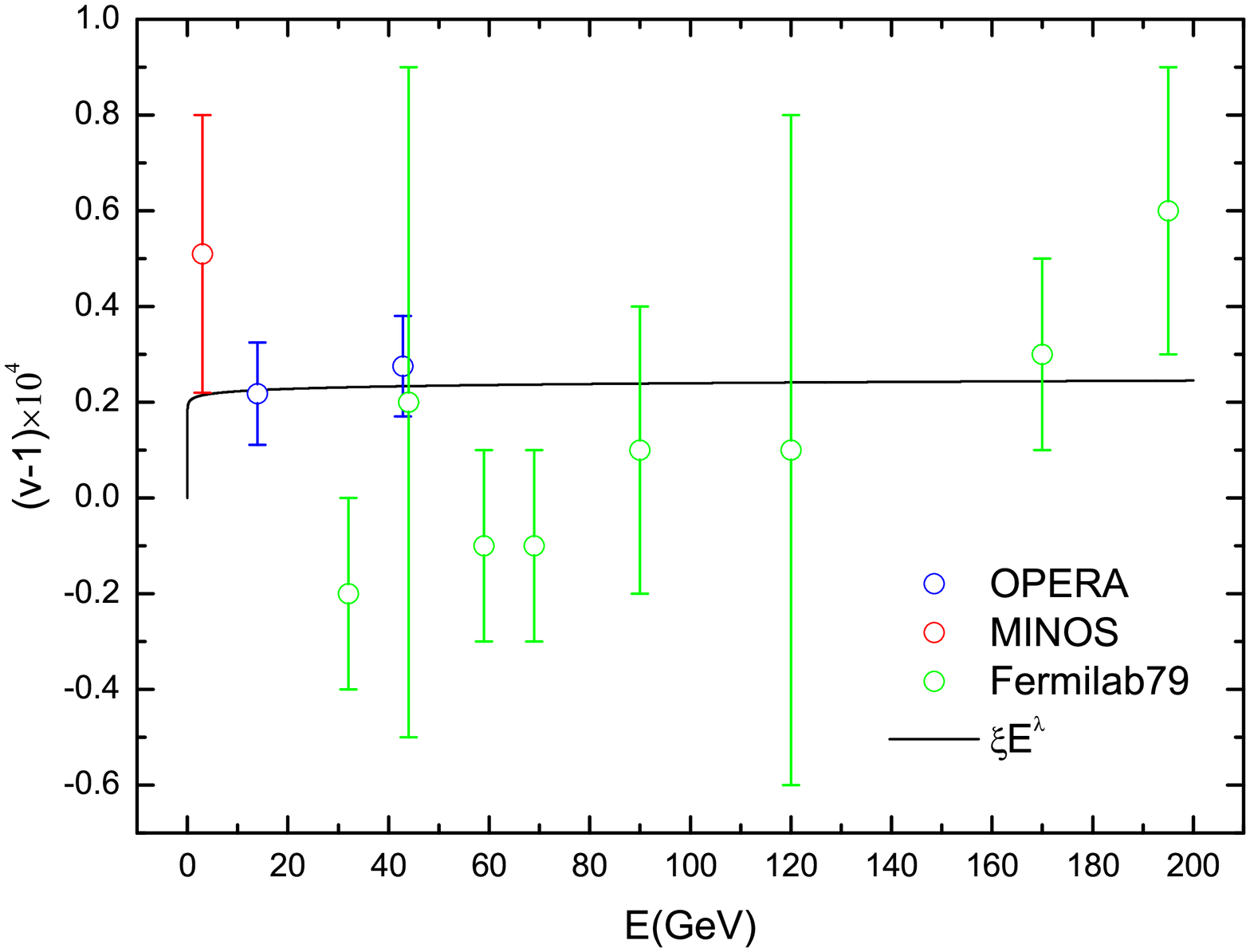} \\
%\mbox{(a)} & \mbox{(b)}
\end{array}$
\end{center}
\caption[]{\small \label{fig:powerlaw}Probability contours at
68.3\%, 95.4\% and 99.7\% confidence level in $\xi-\lambda$ plane
for the power-law scenario, $f(E)=\xi E^\lambda$.}
\end{figure*}

While the CG scenario can fit the superluminal neutrino data well,
it cannot simultaneously explain the result of SN1987A. Can we find
a phenomenological scenario that can reconcile the SN1987A with the
superluminal neutrino experiments such as OPERA, MINOS and
Fermilab79? For this purpose, we test the following power-law
scenario,
\begin{equation}\label{pl}
f(E)=\xi E^\lambda,
\end{equation}
where the energy $E$ is in units of GeV. We now fit to the
experimental data of OPERA, MINOS and Fermilab79 with this model.
From Table~\ref{result} we see that the power-law model yields a
lower $\chi^2_{\rm min}$, 56.78, than that of the CG model. However,
when considering the number of parameters, the power-law model
produces slightly larger BIC and AIC values than the CG model,
namely, $\Delta{\rm BIC}=2.24$ and $\Delta{\rm AIC}=1.84$. This
indicates that the power-law scenario can also fit the data very
well. We show the probability contours in $\xi-\lambda$ plane in the
left panel of Fig.~\ref{fig:powerlaw}. It can be seen from this
figure that the best-fitted $\lambda$ is a tiny positive number, in
the vicinity of zero. In addition, the parameters $\xi$ and
$\lambda$ are in evident degeneracy, strongly anti-correlated. The
right panel of Fig.~\ref{fig:powerlaw} shows the best-fitted $v(E)$
curve of the power-law scenario. We can see that this scenario
behaves just like the CG scenario, and the only difference lies in
the region of energy approaching zero, i.e., in the power-law model
the speed of neutrino will rapidly decrease to the speed of light as
the energy goes to zero, and so the result of SN1987A can also be
successfully accommodated in this scenario.

In summary, we have tested several phenomenological scenarios of
superluminal neutrinos, according to the fits of the experimental
data of OPERA, MINOS and Fermilab79. The purpose of this work is to
see whether or not the speed of the faster-than-light neutrino is
dependent on the energy, from an analysis on the current data. We
adopted a phenomenological perspective that the only focus is on the
data but not the deeper theoretical mechanism. We have shown that
the CG scenario in which the velocity of neutrino is free of the
energy can fit the data best. Nevertheless, this model cannot
meanwhile provide an explanation for the result of SN1987A. We found
that a power-law scenario with the power being a tiny positive
number can simultaneously explain the results of SN1987A and
OPERA+MINOS+Fermilab79. The scenario may deserve a further
investigation.

\begin{acknowledgments}
We thank Nan Li, Yang-Yang Li, Yun-He Li and Jing-Fei Zhang for
useful discussions. This work was supported by the National Science
Foundation of China under Grant Nos.~10705041, 10975032, and
11175042, and by the National Ministry of Education of China under
Grant Nos.~NCET-09-0276 and N100505001.
\end{acknowledgments}
%=================end=======================


\begin{thebibliography}{99}

\bibitem{1109.4897}
%\cite{:2011zb}
%\bibitem{opera:2011zb}
  T.~Adam {\it et al.}  [OPERA Collaboration],
  %``Measurement of the neutrino velocity with the OPERA detector in the CNGS
  %beam,''
  arXiv:1109.4897 [hep-ex].
  %%CITATION = ARXIV:1109.4897;%%

\bibitem{0706.0437}
%\cite{Adamson:2007zzb}
%\bibitem{Adamson:2007zzb}
  P.~Adamson {\it et al.}  [MINOS Collaboration],
  %``Measurement of neutrino velocity with the MINOS detectors and NuMI neutrino
  %beam,''
  Phys.\ Rev.\  D {\bf 76}, 072005 (2007)
  [arXiv:0706.0437 [hep-ex]].
  %%CITATION = PHRVA,D76,072005;%%

\bibitem{1979}
G.~R.~Kalbfleisch {\it et al.}
%``EXPERIMENTAL COMPARISON OF NEUTRINO, ANTI-NEUTRINO, AND MUON VELOCITIES,''
Phys.\ Rev.\ Lett.\  {\bf 43}, 1361 (1979).
%%CITATION = PRLTA,43,1361;%%

%\bibitem{wired.co.uk}
%http://www.wired.co.uk/news/archive/2011-10/17/mundane-neutrino-explanation

\bibitem{gps}
R. A. J. van Elburg, arXiv:1110.2685 [physics.gen-ph].

\bibitem{CG}
%\cite{Cohen:2011hx}
%\bibitem{Cohen:2011hx}
  A.~G.~Cohen and S.~L.~Glashow,
  %``New Constraints on Neutrino Velocities,''
  arXiv:1109.6562 [hep-ph].
  %%CITATION = ARXIV:1109.6562;%%

%\bibitem{1109.5172}
%Giovanni.~{\it et al.}
%OPERA-reassessing data on the energy dependence of the speed of neutrinos
%arXiv:1109.5172

%\cite{AmelinoCamelia:2011dx}
\bibitem{AmelinoCamelia:2011dx}
  G.~Amelino-Camelia, G.~Gubitosi, N.~Loret, F.~Mercati, G.~Rosati and P.~Lipari,
  %``OPERA-reassessing data on the energy dependence of the speed of
  %neutrinos,''
  arXiv:1109.5172 [hep-ph].
  %%CITATION = ARXIV:1109.5172;%%

%\cite{Li:2011ue}
\bibitem{Li:2011ue}
  M.~Li and T.~Wang,
  %``Mass-dependent Lorentz Violation and Neutrino Velocity,''
  arXiv:1109.5924 [hep-ph].
  %%CITATION = ARXIV:1109.5924;%%

%\cite{Lingli:2011yn}
\bibitem{Lingli:2011yn}
  Z.~Lingli and B.~Q.~Ma,
  %``Neutrino speed anomaly as a signal of Lorentz violation,''
  arXiv:1109.6097 [hep-ph].
  %%CITATION = ARXIV:1109.6097;%%

%\cite{Bi:2011nd}
\bibitem{Bi:2011nd}
  X.~J.~Bi, P.~F.~Yin, Z.~H.~Yu and Q.~Yuan,
  %``Constraints and tests of the OPERA superluminal neutrinos,''
  arXiv:1109.6667 [hep-ph].
  %%CITATION = ARXIV:1109.6667;%%

%\cite{Gubser:2011mp}
\bibitem{Gubser:2011mp}
  S.~S.~Gubser,
  %``Superluminal neutrinos and extra dimensions: constraints from the null
  %energy condition,''
  arXiv:1109.5687 [hep-th].
  %%CITATION = ARXIV:1109.5687;%%

%\cite{Ciuffoli:2011ji}
\bibitem{Ciuffoli:2011ji}
  E.~Ciuffoli, J.~Evslin, J.~Liu and X.~M.~Zhang,
  %``OPERA and a Neutrino Dark Energy Model,''
  arXiv:1109.6641 [hep-ph].
  %%CITATION = ARXIV:1109.6641;%%

%\cite{Coleman:1997xq}
\bibitem{Coleman:1997xq}
  S.~R.~Coleman and S.~L.~Glashow,
  %``Cosmic Ray and Neutrino Tests of Special Relativity,''
  Phys.\ Lett.\  B {\bf 405}, 249 (1997)
  [arXiv:hep-ph/9703240].
  %%CITATION = PHLTA,B405,249;%%

%\cite{Coleman:1998ti}
\bibitem{Coleman:1998ti}
  S.~R.~Coleman and S.~L.~Glashow,
  %``High-Energy Tests of Lorentz Invariance,''
  Phys.\ Rev.\  D {\bf 59}, 116008 (1999)
  [arXiv:hep-ph/9812418].
  %%CITATION = PHRVA,D59,116008;%%

%\cite{Liddle:2004nh}
\bibitem{Liddle:2004nh}
  A.~R.~Liddle,
  %``How many cosmological parameters?,''
  Mon.\ Not.\ Roy.\ Astron.\ Soc.\  {\bf 351}, L49 (2004)
  [arXiv:astro-ph/0401198].
  %%CITATION = MNRAA,351,L49;%%

%\cite{Li:2009jx}
\bibitem{Li:2009jx}
  M.~Li, X.~D.~Li and X.~Zhang,
  %``Comparison of dark energy models: A perspective from the latest
  %observational data,''
  Sci.\ China Phys.\ Mech.\ Astron.\  {\bf 53}, 1631 (2010)
  [arXiv:0912.3988 [astro-ph.CO]].
  %%CITATION = 00765,53,1631;%%


\end{thebibliography}
\end{document}